\documentclass[a4paper]{jpconf}
\usepackage{graphicx}
\usepackage{amssymb}

\def\mbh{M_{\rm BH}}
\def\lbol{L_{\rm bol}}
\def\sig{\sigma_{\ast}}
\def\kms{\,{\rm km\,s^{-1}}} 
\def\msun{\,{\rm M}_{\odot}}

\def\ergs{\,{\rm erg\,s}^{-1}}
\def\esc{\,{\rm erg\,s^{-1}\,cm^{-2}}}
\def\mum{\,\mu{\rm m}}
\begin{document}
\title{Discovery of an Intermediate Mass Black Hole at 
the center of the starburst/Seyfert composite galaxy IRAS~01072+4954}

\author{M~Valencia-S.$^{1,2}$, A~Eckart$^{1,2}$, J~Zuther$^1$, S~Fischer$^1$, 
S~Smajic$^{1,2}$, C~Iserlohe$^1$, M~Garc{\'\i}a-Mar{\'\i}n$^1$, L~Moser$^1$, M~Bremer$^{1}$ and M~Vitale$^{1,2}$}

\address{$^1$ I.Physikalisches Institut, Universit\"at zu K\"oln, Z\"ulpicher Str.77, 50937 K\"oln, Germany}
\address{$^2$ Max-Planck-Institut f\"ur Radioastronomie, Auf dem H\"ugel 69, 53121 Bonn, Germany}

\ead{mvalencias@ph1.uni-koeln.de}

\begin{abstract}
The starburst/Seyfert composite galaxy IRAS~01072+4954 ($z=0.0236$) is an enigmatic 
source that combines a Seyfert~1-like X-ray emission with a starburst optical spectrum that
lacks broad line emission. We performed high angular resolution observations of the 
central kiloparsec of this galaxy in the near-infrared. Combining our data with 2MASS 
images of the whole galaxy, we obtain and model the surface brightness profile. We
find indications for the presence of an elongated bar-like structure in both data sets. 
We also model the line of sight velocity distribution of the stars in the bulge.
The derived photometrical and kinematical parameters of the bulge are used to evaluate
the black hole mass through scaling relations. We find that all reliable estimations 
of the black hole mass are consistent with the presence of an intermediate mass
black hole of $\mbh \lesssim 10^5\msun$.    
\end{abstract}

\section{Introduction}
Tight correlations between the mass of supermassive black holes and the properties of their host galaxies \cite{Magorrian1998, Ferrarese2000, Gebhardt2000, Graham2001, MarconiHunt2003} suggest a co-evolution triggered by merging events. At the lower end of the black hole mass range, $\mbh < 10^{7}\msun$, where dwarf galaxies and bulges of late type spirals -- among other sources -- are located, such correlations are less clear \cite{Hu2008, Graham2008, Greene2010, KormendyBender2011, Graham2011pseudo, Jiang2011}. For example, pseudobulges, barred galaxies and Narrow Line Seyfert 1s (NLSy1s) appear to lie below the $\mbh - \sig$ relation established for higher mass systems \cite{Greene2008, Zhou2006, Xiao2011}, which has been interpreted as a sign of evolution triggered by secular processes and/or not fully grown systems \cite{KormendyKennicutt04, Ryan07,  Orban2011, Mathur2011}. Intermediate mass black holes at the center of galaxies (IMBHs; $\mbh \lesssim 10^6\msun$) constitute ideal cases for tracing the behavior of these correlations at the low-mass end. The characteristics of this population are of particular interest for studying galaxy evolution and to put constraints on models of
primordial black hole seed formation \cite{MadauRees2001,Volonteri2003, VolonteriNatarajan2009}.

During the past decade, IMBHs and their hosts have been studied in the optical -- mainly from {\it SDSS} data, but recently also with images taken with WFPC2 on {\it HST} -- and in the X-rays \cite{GreeneHo2005, GreeneHo2007, Dewangan2008, Ai2011, Xiao2011}.  We make use of the high angular and spectral resolution that can be achieved in the near-infrared (NIR) when assisted by adaptive optics systems, to discover an IMBH candidate at the center of the starburst/Seyfert composite galaxy IRAS~01072+4954 ($z = 0.0236$). In the optical, this source displays only narrow emission lines and is classified as a transition object in the Baldwin-Phillips-Terlevich diagnostic diagram \cite{Moran96}. This means that the main ionizing source of the emitting species could be star  formation, an active nucleus (Seyfert 2) or both. However, in the X-ray domain, it shows an emission typical of Seyfert 1 AGN: steep power-law photon-spectral index ($\Gamma=2.1$) and very low hydrogen column density ($N_H < 0.04 \times 10^{22}\,{\rm cm}^{-2}$). Long- and short-term X-ray flux variations were also detected \cite{Panessa05}. 

In our analysis of its nuclear NIR emission ($r \approx 75$\, pc), we find that the main reason for the ambiguous classification is the faintness of the AGN \cite{me12}. The low bolometric luminosity, $\lbol \simeq 2.5 \times 10^{42}\ergs$, combined with a very low $\mbh$ produces broad emission lines that are very faint and narrow, and therefore not detectable in the optical spectrum taken by Moran et al. \cite{Moran96} from the central $(1\times2)\,{\rm kpc^2}$ region.  The predicted flux and width of the H$\alpha$ broad component are $F \simeq 5 \times 10^{-14}\esc$ and {\it FWHM}$\sim (400-600)\kms$ (determined using the expressions from \cite{SternLaor12}; see \cite{me12} for details). Nevertheless, this mini-AGN seems to accrete at a high rate, $\lbol/L_{\rm Edd} \simeq 0.2$, which makes it comparable to other NLSy1s (e.g., 
\cite{Zhou2006}).
 
In this paper we focus on the black hole mass determination using K-band integral field observations of the bulge of this galaxy.  In the following Sect.\,\ref{sec:obs}, we briefly present the data. In Sect.\,\ref{sec:phot}, we show the construction and fitting of the surface brightness profile and discuss the probable presence of an inner structure -- disk or bar. The fitting of the line-of-sight velocity distribution of the stars in the bulge is presented in Sect.\,\ref{sec:kyn}. Finally, in Sect.\,\ref{sec:mbh} we use those photometric and kinematical measurements to estimate the black hole mass via scaling relations and discuss the results. Conclusions are presented in Sect.\,\ref{sec:fin}.
 
\section{The observations}
\label{sec:obs}
The central $\sim 1\,{\rm kpc^2}$ of IRAS~01072+4954 was observed on October 2008 with the Near-Infrared Integral Field Spectrometer NIFS \cite{nifs03} mounted on the ``Frederick C. Gillett'' Gemini North telescope on Mauna Kea, Hawaii. The adaptative optics (AO) module ALTAIR was used in Laser Guide Star mode.
NIFS has a $3''\times 3''$ field of view and provides high spatial and spectral resolution simultaneously. The observations cover the H- and K-bands, although in this paper we refer only to the K-band data. The spatial resolution achieved with the AO in the K-band was $0.15''$, which corresponds to 72\,pc on source. The instrumental spectral resolution is  $57\kms$.  The data was reduced using the GEMINI IRAF package (released Version 2.14, of September 15, 2008) and standard IRAF tasks. Absolute flux calibration and extinction corrections were also applied to the data cubes.

\section{K-band photometry}
\label{sec:phot}

\subsection{Surface brightness profile}
\begin{figure}[t] 
\includegraphics[width=0.58\textwidth]{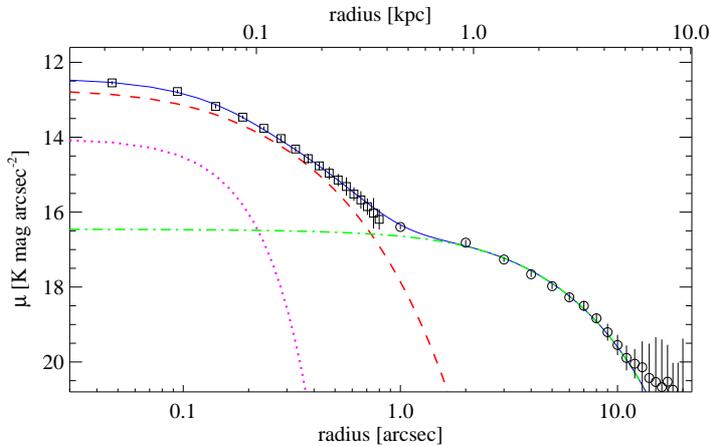}\hspace{2pc}%
\begin{minipage}[b]{0.35\textwidth}\caption{\label{fig:prof} 
K-band brightness profile of IRAS~01072+4954 obtained by the combination of the NIFS (squares) and 2MASS (circles) data. The dashed red line corresponds to a Sersic profile fitted to the bulge, and the dash-dotted green line to an exponential function that fits the disk component. The dotted magenta line represents the AGN + hot dust contribution, while the solid blue line show the overall fit. 
}
\end{minipage}
\end{figure}

The 2MASS \cite{2mass} provides NIR images that cover the whole galaxy. However, in the K-band, the source is very faint -- at the center only $\sim4-5$ times brighter than the background and has a small radius,$\sim 12''$, so that a proper 2D photometric modeling is not possible. We also notice that the bulge size is about the same as the 2MASS K-band point-spread function (PSF), meaning that it is not resolved in the 2MASS image. Therefore, we combine the 2MASS with our NIFS data to construct a 1D surface brightness profile (Fig.\,\ref{fig:prof}). We model the bulge as a Sersic profile with an effective radius $r_e$ and index $n$. The disk is described by an exponential profile with scale radius $h$. A Gaussian with the width of the NIFS PSF represents the emission from the AGN and the hot dust at the center. The photometric parameters resulting from the best fit of these three components are: $r_e=(159.2 \pm 17.8)\,{\rm pc}$, $n = 1.17\pm 0.08$ and $h=(1.47 \pm 0.53)\,{\rm kpc}$ with central surface brightness $I_0 = (74.6 \pm 2.0)\,{\rm L_{\odot,K} \, pc^{-2}}$. 

The presented model is only a first approximation to the description of the surface brightness. There are two main reasons foo believing that the bulge contribution found in this way is overestimated. First, the two data sets do not properly overlap with each other. Assuming that the NIFS data represent the bulge of the galaxy reasonably well, and that the emission from $r \gtrsim 2.5''$ of the 2MASS image corresponds to the disk, we convolve the Sersic model to the resolution of the NIFS data and the exponential function to the one of the 2MASS image. However, these components are not independent of each other and the disk contribution at the center might be underestimated. The second one is that IRAS~01072+4954 is a barred galaxy, and although the bar does not seem to dominate the emission at the center -- as we show in the following -- , its contribution has not been taken into account in the axially symmetric averaged 1D profile.   
			
\subsection{IRAS~01072+4954 bar: Photometrical evidence}

At first glance, the NIR surface brightness of IRAS~01072+4954 in the 2MASS and in the NIFS data look smooth and uniform. In order to recognize the presence of sub-structures in the disk or in the bulge, we use the unsharp masking method. It essentially consists of subtracting a smoothed image of the galaxy from the original one, and  adjusting the smoothing function to strengthen the compact structures. 

\begin{figure}[t]
 \begin{center}
   \includegraphics[width=0.75\textwidth]{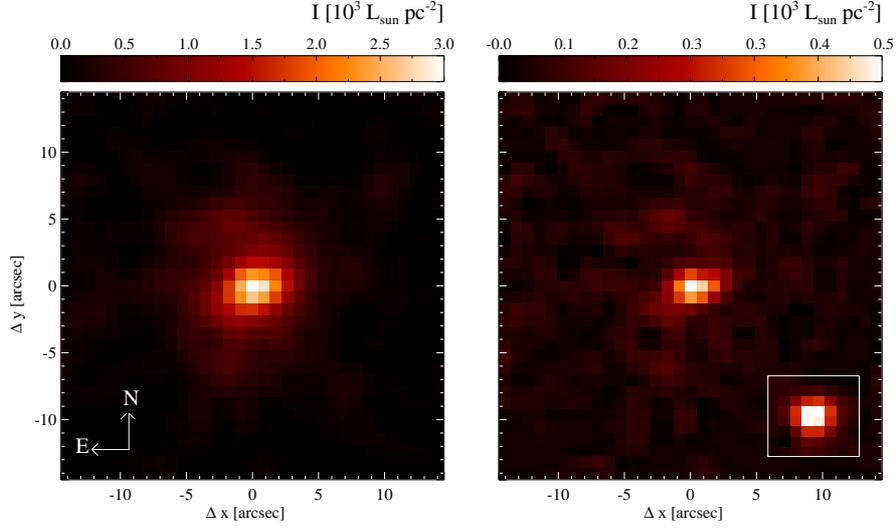}
 \end{center}
\caption{\label{fig:bar2mass}
{\it Left panel:} $15''\times15''$ K-band surface brightness distribution of IRAS~01072+4954 from the 2MASS image. {\it Right panel:} Result of the unsharp masking procedure applied to the image. A star in the field that represents the 2MASS PSF is shown at the bottom right.}
\end{figure}

Figure\,\ref{fig:bar2mass} shows in the second panel the result of this process applied to the $15''\times 15''$ 2MASS K-band image (first panel). The subtracted image was convolved with a 2D Gaussian with a width of  $\sigma=1.5''$. An elongated structure, that resembles a bar, emerges at the center. To evaluate how significant this elongation is, we compare its shape with the 2MASS PSF (image at the bottom right corner on the second panel). Fitting elliptical 2D Gaussians to both shapes, we obtain  axis ratios ($a/b$) of 1.00 for the PSF, and 1.68 with orientation ${\rm P.A.}=105^{\circ}$ for the central structure. This seems to indicate that the shape is intrinsic to the sub-structure. However, the sizes are comparable, implying that the bar-like feature is not resolved in the 2MASS image. 
 
\begin{figure}[ht]
 \begin{center}
   \includegraphics[width=0.75\textwidth]{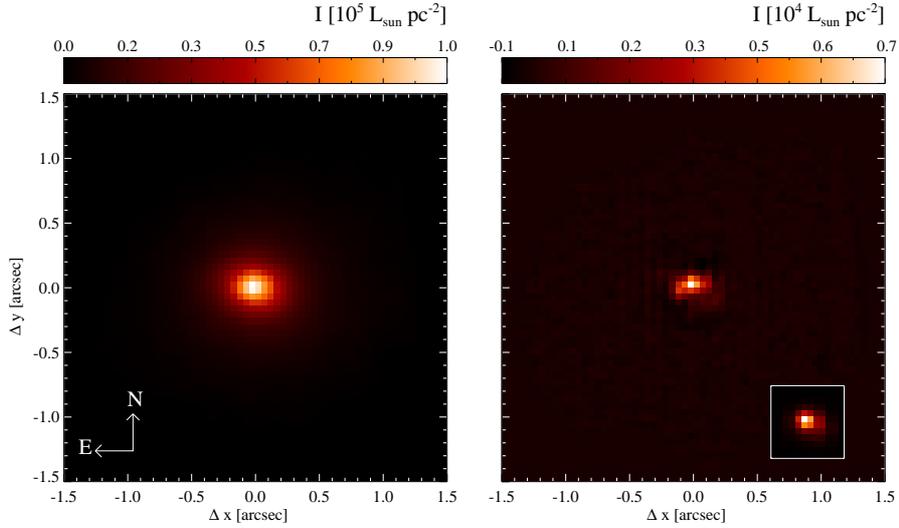}
 \end{center}  
\caption{\label{fig:barnifs}
Same as in Fig.\,\ref{fig:bar2mass} but for the $3''\times3''$ NIFS image. The residuals in the right panel are obtained after the removal of the strong PSF contribution (see text for details).}
\end{figure}

When the same procedure is applied to the $3''\times3''$ NIFS image, the AGN and hot dust emission at the center dominate the residuals. Their contribution corresponds to the PSF, which is assumed to be well described by the image of a star close to the field of view. To remove this emission, the PSF is subtracted from the K-band continuum image by the maximum amount possible such that the residual image is still smooth at the center. Then, the unsharp masking procedure is applied convolving the image with a Gaussian of $\sigma\approx0.07''$.
The second panel in figure\,\ref{fig:barnifs} shows the results. In the first panel the K-band continuum image, created by taking a median of the NIFS data cube over the wavelength range, is shown. The unsharp masking method reveals an elongated structure. The faint and more extended emission at the center probably corresponds to residuals of the PSF subtraction. When fitting the elongated feature with a 2D Gaussian, we find that the axis ratio $a/b=1.74$ and the orientation ${\rm P.A.}=97^{\circ}$ are comparable to the values found previously with the 2MASS data. This allows us to suspect that both residuals are related to the same sub-structure. In this case, the sensitivity of the observations is not high enough to obtain a better map of it, and for the same reason, cannot be resolved. As shown in Fig.~\,\ref{fig:barnifs}, its surface brightness is on average $\sim 3.0\times10^3 \,{\rm L_{\odot,K} \, pc^{-2}}$. This is the same brightness as the one observed in the continuum image at about 1\,arcsecond from the center. Although it is unlikely that its contribution would change by much more than about one order of magnitude our estimation of the bulge luminosity, an exact estimation is not possible because we are missing all emission fainter than $\sim 16\,{\rm mag\,arcsec^{-2}}$ -- which includes the disk -- in the NIFS observations.

\section{Stellar kinematics}
\label{sec:kyn}

\subsection{Velocity distribution}
We derive the line-of-sight velocity distribution (LOSVD) of the stars in the photometric bulge, by fitting a set of high angular resolution stellar templates to the spectrum integrated over an aperture of $r\approx0.34''$, which corresponds to the scale radius of the bulge found above. We assume that the stellar LOSVD can be described by a truncated Gauss-Hermite series parametrized by the radial velocity $V$, the velocity dispersion $\sigma_{\ast}$, and coefficients of the third and fourth Hermite polynomials $h_3$ and $h_4$. The deep and narrow CO absorption features at $2.294\mum$ and $2.323\mum$ indicate that the velocity dispersion of the stars in the bulge is low. This implies that, despite the high spectral resolution of the NIFS data, it is not possible to constrain significantly all the LOSVD parameters simultaneously \cite{Cappellari04}. In this cases, one would like to bias the solution towards the pure Gaussian one and, in that way, avoid the coupling of the parameters, $V$ with $h_3$ and $\sigma_{\ast}$ with $h_4$. The penalized pixel fitting routine (pPXF) of Cappellari and Emsellem \cite{Cappellari04} applies such procedure, allowing to recover the LOSVD in moderately low signal-to noise ($S/N$) and/or undersampled data. This routine minimizes the difference between the galaxy spectrum and the combination of the stellar templates convolved with the LOSVD. The algorithm allows different weights of the stellar templates. We use a subset of 
the Gemini spectroscopic library of NIR  stars observed with the NIFS IFU in K-band \cite{Winge09}. To reduce the effects of the template mismatch, we select the stellar types that match best the absorption features in both H and K bands by comparing the galaxy spectrum with medium resolution stellar templates of Ivanov et al. \cite{Ivanov04}. The selected set of stellar spectra consists of 7 objects of the following types: K2III, K5Ib,K5II,K5III, M2III, M3III and M5III. 

\begin{figure}[h] 
\includegraphics[width=0.58\textwidth]{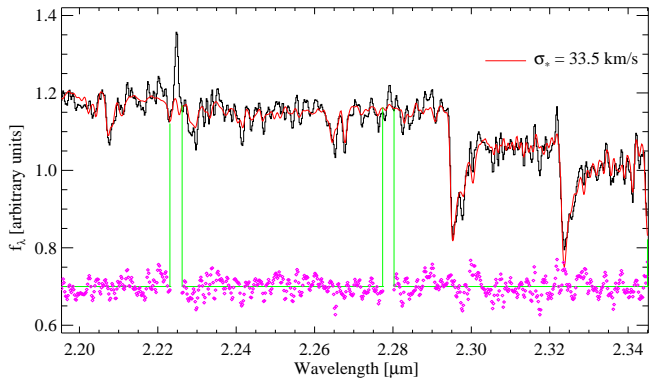}\hspace{2pc}%
\begin{minipage}[b]{0.35\textwidth}\caption{\label{fig:coppxf} 
Fit of the stellar kinematics of the spectrum integrated over an aperture $r=r_e=160\,{\rm pc}$. The observed spectrum is shown as a thin solid line, while the red solid line represents the best fit to the LOSVD.  The small spectral regions delimited by the vertical green lines correspond to two H$_2$ emission lines, which were masked for the fit. The residuals, shown at the bottom, are shifted vertically for clarity. 
}
\end{minipage}
\end{figure}
 
Figure\,\ref{fig:coppxf} shows the best fit of the LOSVD with the parameters: $V=(-17.1\pm 1.6)\kms$, $\sigma_{\ast}=(33.5 \pm 3.7)\kms$, $h_3=0.06\pm0.05$ and $h_4=0.01\pm0.08$. The fit is obtained after adopting a ${\rm bias}=0.5$ and the addition of a Legendre polynomial of degree 10. In a conservative approach, the errors are calculated via Monte Carlo simulations, where 1000 realizations of an input spectrum are fitted with the pPXF routine without any penalization (${\rm bias}=0.0$). The input spectrum is created by convolving a stellar template with the LOSVD and adding Poisson noise to the level of the galaxy spectrum, in this case, to reach $S/N\sim40$. Then, the effects of undersampling are simulated allowing the velocity scale ($\kms$ per pixel) of the input spectrum to vary in such a way that the second moment of the LOSVD, found by fitting the spectrum with the pPXF routine, varies in the range $\sim14-140\kms$. Given the moderate $S/N$ of the data and that $\sigma_{\ast}$ is low compared to the velocity scale (less than two pixels), the higher moments $h_3$ and $h_4$ of the LOSVD suffer from large uncertainties. Fitting only the first two moments we obtain $V=-16.1\kms$ and  $\sigma_{\ast}=32.9\kms$, which are within the estimated errors. 

\subsection{IRAS~01072+4954 bar: Kinematical evidence}

\begin{figure}[h]
\includegraphics[width=\textwidth]{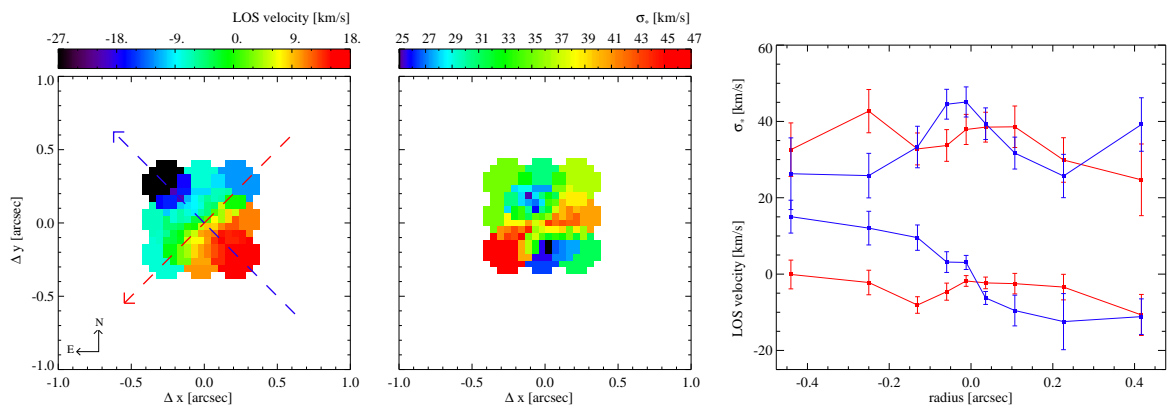}
\caption{\label{fig:losvd}
{\it Left panel:} LOS velocity. {\it Middle panel:} Velocity dispersion. {\it Right panel:} LOS velocity and velocity dispersion curves obtained from simulated long-slit spectroscopy along the two axes shown in the left panel. Blue curves+points correspond to the slit along the axis with higher LOS velocity gradient, red curves+points to a slit perpendicular to the previous one. The radius to the center increases from west to east. (These are preliminary results).}
\end{figure}

Though the bar-like structure is barely detected photometrically, the kinematics of the stars very close to the center are clearly affected by its presence. To elaborate a map of the stellar LOSVD, we bin the data in apertures with three different sizes that have $S/N > 15$ per bin. Then, we fit the LOSVD of the spectra integrated in every aperture using the pPXF routine, and combine the results. This procedure gives us a first estimate of the stellar velocity field (Fig.\,\ref{fig:losvd}). Special care has to be taken when interpreting the maps because the combination procedure smooths the velocity gradients, and the error of the values increases with the distance to the center. Given the low values of LOS velocity and velocity dispersion, the $S/N$ of single pixel spectra, even after binning the data only information of the central square arcsecond ($\sim 460\,{\rm pc}$) can be recovered.    

First and second panels in Fig.\,\ref{fig:losvd} show the stellar LOS velocity and the velocity dispersion maps constructed from the NIFS K-band data cube. The first one displays a characteristic rotation pattern consistent with the expectations for a disk galaxy nearly face-on. In the dispersion map, there is a clear evidence for an internal structure with ${\rm P.A.}\approx 120^{\circ}$.  To get a better estimation of the rotation curve and the uncertainties, we simulate long slits of $\sim1$\,arcsec width that cross the center with different orientations. The data of every slit are binned to have the maximum number of points with $S/N$ in the range 10-20. Finally, the pPXF routine is applied tuning the bias in every single bin to obtain the best fit and the uncertainties in each case. The third panel in Fig.\,\ref{fig:losvd} shows the derived rotation and dispersion curves for the slit oriented along the highest LOS velocity gradient (${\rm P.A.}\approx 45^{\circ}$) and those for the slit perpendicular to it.  

The fact that the velocity dispersion is dominated by the bar-like structure implies that the $\sigma_{\ast}$ estimated above is an upper limit. 


\section{Discussion: Black hole mass estimations}
\label{sec:mbh}
The mass of the black hole at the center of galaxies has been found to correlate with some properties of the host galaxy bulge.  However, barred galaxies show a higher scatter around the scaling relations defined by normal disk galaxies or follow a different relation \cite{Hu2008, Graham2008}. It seems that the vertical resonances and/or instabilities responsible for the formation of bars can affect dramatically the velocity distribution of the stars, and favor the development of pseudobulges \cite{KormendyKennicutt04, Athanassoula2005, Graham2011}. Here we use the derived upper limits of the stellar velocity dispersion and the K-band bulge luminosity to estimate the black hole mass $\mbh$. The validity of the scaling relations used is also discussed. Table\,\ref{table:mbh} summarizes the different black hole mass estimations.

\subsection{$\mbh - L_{\mathrm{K}}$}
The K-band bulge luminosity, obtained by integrating the Sersic profile that best fit the surface brightness, is $L_{\mathrm{K}}=3.24\times10^{9}\,{\rm L_{K,\odot}}$. As explained previously, we expect this value to be overestimated by $\sim 1$ order of magnitude. In \cite{me12} we show that using the $\mbh - L_{\mathrm{K}}$ relation from Marconi \& Hunt \cite{MarconiHunt2003}, the upper limit on the black hole mass is $2.5\times10^6\msun$, while with the Graham updated version of the same relation \cite{Graham2007} it is $9.8\times10^6\msun$. However, as Graham pointed out (private communication, also \cite{Graham2012}), such a relation is not suitable for low-$\mbh$, because it has been derived mainly based on luminous core galaxies. Core galaxies follow a luminosity-dispersion $L \propto \sigma_{\ast}^5$ relation and their $\mbh \propto L^1$, probably as a result of a dry merging formation process \cite{Graham2012} (see also \cite{Kormendy2009}). In contrast, non-barred low-$\mbh$ Sersic galaxies follow a $\mbh \propto L^{2.5}$ relation. Given the lack of a study that investigates the $\mbh - L_{\mathrm{K}}$ in such systems, we make the exercise to bend the known $\mbh - L_{\mathrm{K}}$ relation (Eq.14 of \cite{Graham2007}) at $\sim 7\times10^7\msun$ assuming the stated proportionality. With the resulting relation, $\log(\mbh/\msun)\sim2.5\log(L_{\mathrm{K}}/{\rm L_{K,\odot}})-18.23$, we  obtain a black hole mass of $3.5\times10^5\msun$. 

\subsection{$\mbh - \sigma_{\ast}$}  
Recently G\"{u}ltekin et al. \cite{Gultekin2009} and Graham et al. \cite{Graham2011} have recalibrated the $\mbh - \sigma_{\ast}$. They use 49 objects and 18 upper limits, and 64 objects, respectively, with reliable parameter estimations. Also Xiao et al. \cite{Xiao2011} studied this relation, focusing on low-$\mbh$ sources. We evaluate the black hole mass using their relations and report the results in Table\,\ref{table:mbh}. 

\begin{table}[h]
\caption{\label{table:mbh} Black hole mass estimations from scaling relations.} 
\begin{center}
\lineup
\begin{tabular}{clc}
\br                              
$\log \left( \frac{\mbh}{\msun} \right)$ & \0\0\0\0\0\0\0\0\0\0\0\0notes on the applied scaling relation & ref.\cr 
\mr
\multicolumn{3}{c}{$\mbh - L_{\mathrm{K}}$}\cr 
\mr
$6.40$ & 27 bright, mainly early type objects. Not suitable in the case (see text). &\cite{MarconiHunt2003}\cr
$7.00$ & 22 bright objects. Not suitable in the case (see text).&  \cite{Graham2007}\cr 
$5.54$ & Bending the Eq.14 of \cite{Graham2007}. See text for details. &---\cr
\mr
\multicolumn{3}{c}{$\mbh - \sigma_{\ast}$}\cr 
\mr
$4.83$ & Full sample. Intrinsic scatter 0.44 dex. & \cite{Gultekin2009}\cr 
$6.83$ & 8 + 11 upper limits barred galaxies. Might not be reliable (see text).  & \cite{Gultekin2009}\cr 
$3.53$ & Full sample. Intrinsic scatter 0.35 dex. & \cite{Graham2011}\cr 
$3.75$ & 20 barred galaxies. Intrinsic scatter 0.31 & \cite{Graham2011}\cr
$5.10$ & 155 low-$\mbh$ sources. Intrinsic scatter 0.46 & \cite{Xiao2011}\cr 
$4.61$ & 25 low-$\mbh$ barred galaxies. Intrinsic scatter 0.46 & \cite{Xiao2011}\cr 
\br
\end{tabular}
\end{center}
\end{table}

From the black hole masses obtained through the $\mbh - \sigma_{\ast}$ relation, the highest value corresponds to the one calculated from the relation found by G\"{u}ltekin et al. for barred galaxies. However, when looking at the location of these sources in the $\mbh$ vs. $\sigma_{\ast}$ diagram (their Fig~1.), the 8 galaxies with dynamically detected black holes lie well within the scatter of the correlation depicted by the full sample (49 sources). Nevertheless, the $\mbh$ calculated with the relation fitted to the whole sample  is two orders of magnitude lower than the one obtained by using only barred galaxies (+ 18 upper limits). The authors also assign a probability of 0.1809 to the case of the upper limit sources not having a black hole. In contrast, the same probability stated for the full sample is 0.0004. We conclude that the difference between the two estimations is introduced by the inclusion of the upper limits to the fit which, in the case of the barred galaxies, dominate the subsample. Therefore, we do not consider this black hole mass estimation as reliable. All other estimations of the $\mbh$ are consistent with the hypothesis of the existence of an IMBH at the center of IRAS~01072+4954. In any case, it is interesting to notice the differences in the $\mbh$ calculated from various versions of the scaling relations. This of course is related to the sample selection, the method applied to fit the data, and the treatment of the errors.   

The question of whether the emission detected in the NIFS data corresponds to a classical bulge or to a pseudobulge is still open. High angular resolution observations of the whole galaxy or a major portion of it are necessary to do a proper decomposition of the photometrical components and derive more accurate bulge parameters.

\section{Summary and conclusions}
\label{sec:fin}

IRAS~01072+4954 is a starburst/Seyfert galaxy that hosts an active nucleus detected in the NIR and in X-rays \cite{me12, Panessa05}. Here we expose the methods employed to get photometrical and kinematical information of the bulge, a more detailed description will be published in a forthcoming paper. 

We present evidence for the presence of an IMBH at the center of IRAS~01072+4954. We make use of scaling relations of the black hole mass with the K-band bulge luminosity and with the stellar velocity dispersion recently calibrated with different samples. All the reliable $\mbh$ estimations derived from those relations are consistent with $\mbh \lesssim 10^5\msun$.

We also find indications for the presence of a bar-like structure with ${\rm P.A.}\sim 100^{\circ}$. This implies an overestimation of the bulge luminosity, the velocity dispersion and, consequently, of the black hole mass.

\ack
M. Valencia-S thanks J. Stern and A. Laor for constructive discussions during the 
AHAR Conference, and also A. Graham for his comments, in particular referring to the 
suitability of the scaling relations to derive black hole masses. 
M. Valencia-S., S. Smajic and M. Vitale are members of the International 
Max-Planck Research School (IMPRS) for Astronomy and Astrophysics at the 
Universities of Bonn and Cologne supported by the Max Planck Society. 
J. Zuther, S. Fischer and M. Bremer gladly acknowledge the financial support 
by Group of Eight Australia-Germany Joint Research Co-operation Scheme via 
project-ID 50753527. 
M. Garc\'{\i}a-Mar\'{\i}n is supported by the German federal department for
education and research (BMBF) under the project No. 50OS1101.
Part of this work was supported by the German Deutsche 
Forschungsgemeinschaft, DFG, via grant SFB 956 and fruitful discussions with 
members of the European Union funded COST Action MP0905: Black Holes in a 
violent Universe and PECS project No. 98040. 
Based on observations  (Programm ID: GN-2008B-Q-77) obtained at the Gemini 
Observatory, which is operated 
by the Association of Universities for Research in Astronomy, Inc., under 
a cooperative agreement with the NSF on behalf of the Gemini partnership: 
the National Science Foundation (United States), the Science and Technology 
Facilities Council (United Kingdom), the National Research Council (Canada), 
CONICYT (Chile), the Australian Research Council (Australia), Minist\'{e}rio 
da Ci\^{e}ncia, Tecnologia e Inova\c{c}\~{a}o (Brazil) and Ministerio de 
Ciencia, Tecnolog\'{i}a e Innovaci\'{o}n Productiva (Argentina).

\end{document}